\global\def\draftcontrol{0}

   \def\versionno{ holomorphic anomaly }
\catcode`\@=11

\expandafter\ifx\csname draftcontrol\endcsname\relax\global\def\draftcontrol{0} 
\fi 

{\count255=\time\divide\count255 by 60 
\xdef\hourmin{\number\count255} 
\multiply\count255 by-60\advance\count255 by\time 
\xdef\hourmin{\hourmin:\ifnum\count255<10 0\fi\the\count255}} 
\def\draftdate{\number\month/\number\day/\number\year\ \ \ \hourmin } 

\newcommand\makepapertitle{\par

  \begingroup 
    \renewcommand\thefootnote{\@fnsymbol\c@footnote}% 
    \def\@makefnmark{\rlap{\@textsuperscript{\normalfont\@thefnmark}}}% 
    \long\def\@makefntext##1{\parindent 1em\noindent 
            \hb@xt@1.8em{% 
                \hss\@textsuperscript{\normalfont\@thefnmark}}##1}% 
     \newpage 
     \global\@topnum\z@   % Prevents figures from going at top of page. 
     \@makepapertitle 
     \thispagestyle{empty}\@thanks 
  \endgroup 
  \setcounter{footnote}{0}% 
  \global\let\thanks\relax 
  \global\let\makepapertitle\relax 
  \global\let\@makepapertitle\relax 
  \global\let\@thanks\@empty 
  \global\let\@author\@empty 
  \global\let\@date\@empty 
  \global\let\@title\@empty 
  \global\let\title\relax 
  \global\let\author\relax 
  \global\let\date\relax 
  \global\let\and\relax 
  \def\version{\let\version\@version\@gobble} 
} 
\def\@makepapertitle{% 
  \newpage 
   \ifnum\draftcontrol=1 {} 
   \version\versionno 
   \vskip 5em% 
   \else 
   \hfill\hbox to 3cm {\parbox{4cm}{\@pubnum}\hss}% 
   \vskip 5em% 
   \fi 
   \begin{center}% 
   \let \footnote \thanks 
      {\hskip -0\textwidth \hbox to 1\textwidth% 
        {\centerline{\Large\bf{\noindent\@title}}}}% 
     \vskip 1.5em% 
     {\normalsize%\large 
       \lineskip .5em% 
       \begin{tabular}[t]{c}% 
         \@author 
       \end{tabular}\par}% 
     \vskip 1.5em% 
     {\@bstract}% 
     \end{center}% 
     \vfill
     \@date%
     \vskip 1.5em%
   \par 
} 

\gdef\@pubnum{} 
\def\pubnum#1{% 
  \gdef\@pubnum{#1}} 

\gdef\@bstract{} 
\def\Abstract#1{% 
  \gdef\@bstract{% 
   \parbox{\textwidth-0pc}{% 
   \centerline{\bf Abstract}\penalty1000 
   \noindent%\abstractfont \baselineskip=12pt 
   \renewcommand\baselinestretch{1.0} 
   {#1}}} 
} 

\gdef\@email{}
\def\email#1{%
   \gdef\@email{%
   Email: {\tt #1}}
}

\def\ps@paper{\let\@mkboth\@gobbletwo% 
     \ifnum\draftcontrol=1 
        \def\@oddfoot{\hbox to \textwidth{\tiny \versionno \hfil\tiny\draftdate}% 
        \hskip -\textwidth \hbox to \textwidth{\hfil\rm\thepage\hfil}}% 
     \else\def\@oddfoot{\hbox to \textwidth{\hfil\rm\thepage\hfil}} 
     \fi 
     \let\@evenfoot\@oddfoot 
} 
\def\body{\clearpage 
          \pagestyle{paper} 
        } 
\newenvironment{acknowledgments}{% 
\vskip 3.25ex 
\addcontentsline{toc}{section}{Acknowledgments}
\noindent {\bf Acknowledgments} 
} 

\def\@version#1{\ifnum\draftcontrol=1 
\typeout{}\typeout{#1}\typeout{} 
\vskip3mm\centerline{\hbox{\fbox{\normalsize{\tt DRAFT -- #1 -- } 
                   {\draftdate}}}}\vskip3mm 
\fi} 
\let\version\@version 
\long\def\eqlabel#1{\ifnum\draftcontrol=1 
                    \tag@false  % there are some problems with multline without this 
                    \tag*{(\theequation) \hbox to -0.2cm{\hspace{0cm}\small{#1}\hss}} 
                    \refstepcounter{equation}  
                    \edef\@currentlabel{\theequation} 
                    \ltx@label{#1}          % use old LaTeX \label instead of new definition 
                    \else 
                    \label{#1} 
                    \fi 
                    } 
\let\st@bibitem\@bibitem 
\let\st@lbibitem\@lbibitem 
\ifnum\draftcontrol=1 
  \def\@bibitem#1{% 
    \st@bibitem{#1}\a@@label{#1}\ignorespaces} 
  \def\@lbibitem[#1]#2{% 
    \st@lbibitem[#1]{#2}\a@@label{#2}\ignorespaces} 
  \def\a@@label#1{% 
    \gdef\a@lab{\smash{\normalfont\small#1}} 
    \ifvmode 
      \if@inlabel 
        \global\setbox\@labels\hbox{% 
          \llap{\a@lab\let\a@lab\relax 
                \kern\@totalleftmargin\kern\marginparsep}% 
          \box\@labels}% 
      \fi 
    \fi} 
\fi 
\documentclass[12pt,letterpaper]{article} 
\usepackage{amsmath,bm,amsfonts,amssymb,array,calc,amsthm,rotating}
\usepackage{epsfig,psfrag} 
\usepackage{graphicx}
\usepackage{color}
\usepackage[colorlinks=false]{hyperref}
\renewcommand\baselinestretch{1.25} 
\renewcommand\section{\@startsection {section}{1}{\z@}% 
                                   {-3.5ex \@plus -1ex \@minus -.2ex}% 
                                   {2.3ex \@plus.2ex}% 
                                   {\normalfont\large\bfseries}} 
\renewcommand\subsection{\@startsection{subsection}{2}{\z@}% 
                                   {-3.25ex\@plus -1ex \@minus -.2ex}% 
                                   {1.5ex \@plus .2ex}% 
                                   {\normalfont\normalsize\bfseries}} 
\renewcommand\subsubsection{\@startsection{subsubsection}{3}{\z@}% 
                                   {-3.25ex\@plus -1ex \@minus -.2ex}% 
                                   {1.5ex \@plus .2ex}% 
                                   {\normalfont\normalsize\it}} 
\renewcommand\paragraph{\@startsection{paragraph}{4}{\z@}% 
                                   {-3.25ex\@plus -1ex \@minus -.2ex}% 
                                   {1.5ex \@plus .2ex}% 
                                   {\normalfont\normalsize\bf}} 
\renewcommand\subparagraph{\@startsection{subparagraph}{5}{\z@}% 
                                   {-1.25ex\@plus -1ex \@minus -.2ex}% 
                                   {0ex \@plus .2ex}% 
                                   {\normalfont\normalsize\it}}

\numberwithin{equation}{section}

\long\def\@makecaption#1#2{%
  \vskip\abovecaptionskip
  \sbox\@tempboxa{{\bf #1:} #2}%
  \ifdim \wd\@tempboxa >\hsize
    {\small\bf #1:} {\small #2}\par
  \else
    \global \@minipagefalse
    \hb@xt@\hsize{\hfil\box\@tempboxa\hfil}%
  \fi
  \vskip\belowcaptionskip}
\renewcommand*\l@section[2]{%
  \ifnum \c@tocdepth >\z@
    \addpenalty\@secpenalty
    \addvspace{.5em \@plus\p@}%
    \setlength\@tempdima{1.5em}%
    \begingroup
      \parindent \z@ \rightskip \@pnumwidth
      \parfillskip -\@pnumwidth
      \leavevmode \bfseries
      \advance\leftskip\@tempdima
      \hskip -\leftskip
      #1\nobreak\hfil \nobreak\hb@xt@\@pnumwidth{\hss #2}\par
    \endgroup
  \fi}
\renewcommand*\l@subsection{\addvspace{.0em \@plus\p@}\@dottedtocline{2}{1.5em}{2.3em}}
\renewcommand*\l@subsubsection{\addvspace{-.2em \@plus\p@}\@dottedtocline{3}{3.8em}{3.2em}}

\def\hepth#1{\href{http://xxx.arxiv.org/abs/hep-th/#1}{{arXiv:hep-th/#1}}}

\def\mathph#1{\href{http://xxx.arxiv.org/abs/math-ph/#1}{{arXiv:math-ph/#1}}}

\def\math#1{\href{http://xxx.arxiv.org/abs/math/#1}{{arXiv:math/#1}}}

\def\alggeom#1{\href{http://xxx.arxiv.org/abs/alg-geom/#1}{{arXiv:alg-geom/#1}}}
\def\arxiv#1#2{\href{http://xxx.arxiv.org/abs/#1}{{arXiv:#1 [#2]}}}

\definecolor{refcol}{rgb}{0.2,0.2,0.8}
\definecolor{eqcol}{rgb}{.6,0,0}
\definecolor{purple}{cmyk}{0,1,0,0}

\gdef\@citecolor{refcol}
\gdef\@linkcolor{eqcol}
\def\colorlinkspurple{\gdef\@urlcolor{purple}}
\def\colorlinksblue{\gdef\@urlcolor{blue}}
\def\colorlinksred{\gdef\@urlcolor{red}}

\def\ie{{\it i.e.}} 
\def\eg{{\it e.g.}} 

\def\cf{{\it cf.}}

\def\revise#1       {\raisebox{-0em}{\rule{3pt}{1em}}% 
                     \marginpar{\raisebox{.5em}{\vrule width3pt\ 
                     \vrule width0pt height 0pt depth0.5em 
                     \hbox to 0cm{\hspace{0cm}{% 
                     \parbox[t]{4em}{\raggedright\footnotesize{#1}}}\hss}}}}

\def\calf         {{\cal F}}

\def\calh         {{\cal H}}

\def\call         {{\cal L}} 
\def\calm         {{\cal M}} 
\def\tcalm        {\widetilde{\cal M}} 
\def\caln         {{\cal N}}

\def\calt         {{\cal T}}

\def\complex      {{\mathbb C}}

\def\rationals    {{\mathbb Q}} 
\def\reals        {{\mathbb R}} 
\def\zet          {{\mathbb Z}} 

\def\del          {\partial} 
\def\delbar       {\bar\partial} 
\def\ee           {{\it e}} 
\def\ii           {{\it i}}

\def\Im           {{\rm Im\hskip0.1em}}

\newcommand\topa[2]{\genfrac{}{}{0pt}{2}{\scriptstyle #1}{\scriptstyle #2}}

\def\sqr#1#2{{\vcenter{\vbox{\hrule height.#2pt   
 \hbox{\vrule width.#2pt height#1pt \kern#1pt 
 \vrule width.#2pt}\hrule height.#2pt}}}}

\def\End{{\rm End}}

\def\Im{{\rm Im}}

\def\ib{{\bar i}}
\def\jb{{\bar j}}
\def\kb{{\bar k}}
\def\lb{{\bar l}}

\def\0b{{\bar 0}}

\def\F#1#2{{\calf}^{(#1,#2)}}

\def\Fc#1{{\calf}^{(#1)}}

\def\griff{{\rm Griff}}

\def\Ipp{\mathord{\mathchar "0271 \kern-4.5pt \mathchar"0271}}

\newcommand{\hC}{\hat{C}}
\newcommand{\IP}[1]{\langle#1\rangle}

\catcode`\@=12

\begin{document} 

%%% 
%%%%%% text starts here 
%%%%%%%%% 

\title{
\parbox{\textwidth}{\begin{center}
Background Independence and the Open Topological
String Wavefunction
\end{center} }}

\pubnum{%
arXiv:0709.2390}
\date{September 2007}

\author{
Andrew Neitzke and Johannes Walcher \\[0.2cm]
\it School of Natural Sciences, Institute for Advanced Study\\
\it Princeton, New Jersey, USA
}

\Abstract{
The open topological string partition function in the background of a
D-brane on a Calabi-Yau threefold specifies a state in the Hilbert 
space associated with the quantization of the underlying special 
geometry. This statement is a consequence of the extended 
holomorphic anomaly equation after an appropriate shift of the
closed string variables, and can be viewed as the expression of
background independence for the open-closed topological string.
We also clarify various other aspects of the structure of 
the extended holomorphic anomaly equation. We conjecture that the 
collection of all D-branes furnishes a basis of the Hilbert space, 
and revisit the BPS interpretation of the open topological string 
wavefunction in this light.
}

%\enlargethispage{1.5cm}

\makepapertitle

\body

\version\versionno

\vskip 1em

%\tableofcontents
%\newpage

\section{Introduction}

In this paper, we study topological strings on Calabi-Yau threefolds
with background D-branes. Our most basic motivation is to understand 
the properties of the open topological string amplitudes $\F gh$ that 
are implied by the extended holomorphic anomaly equations recently
found in \cite{extended}. Two of the applications we have in mind
are open-closed duality in the topological string, and the relation of 
the topological string to BPS state counting.

\subsection{Background}

Without attempting a complete history of the subject, we note that
the holomorphic anomaly equations were originally obtained by BCOV 
\cite{bcov1,bcov2} as a kind of generalization of the $tt^*$-equations 
of Cecotti and Vafa \cite{ceva} to higher-genus amplitudes. These 
equations arise as constraints on the amplitudes of certain 
topologically twisted $\caln=2$ superconformal field theories, and
the coupling of the latter to topological gravity. The original
$tt^*$-equations, which apply more generally also to topological twists of
massive theories, have an appealing geometrical interpretation in 
the context of Frobenius manifolds \cite{dubrovin}, and have attracted
much interest over the years. The mathematical scope of 
the holomorphic anomaly equations is somewhat less well understood,
although of course when combined with mirror symmetry, they are a 
powerful tool to access the enumerative geometry of Calabi-Yau 
manifolds, see \cite{hkq,hosono} for the state of the art.
Physically, the holomorphic anomaly equations have an elegant 
interpretation as the realization of ``quantum background
independence'' of the topological string \cite{wittenwf}, also known 
as the ``wavefunction interpretation''. This slightly mysterious notion 
has recently played a central role in relating the topological string 
to BPS state counting \cite{osv}.

One of the recurring themes in the literature on the $tt^*$-equations, 
the topological string, and the holomorphic anomaly equations is the
relation to classical integrable systems and their quantization.
Recent examples include the solution of the topological string 
on certain local Calabi-Yau manifolds in terms of matrix models 
\cite{diva}, especially when viewed in the broader context of 
the duality between open and closed topological strings \cite{adkmv}.

Our present work is concerned with some extensions of 
these structures to the situation with D-branes in the background.
We believe that our results provide further hints of how the various
topics listed above might be related at a deeper level. In particular,
our results reinforce the point of view that D-branes and open-closed
duality are central to a complete understanding of the wavefunction 
interpretation of the topological string, background independence, as 
well as the underlying integrable structure. Although there are notable 
differences from previous works, we attribute most of them to the distinction 
between local and compact setups. For example, the recent works 
\cite{marcos1,eynard} have  shown that topological string 
amplitudes that can be obtained from matrix models satisfy
a set of equations essentially equivalent to the holomorphic anomaly 
equations.\footnote{More precisely, the approach of \cite{marcos1,eynard}, 
which is based on new techniques in \cite{eynardorantin}, will work quite
generally for local Calabi-Yau manifolds which are conic bundles 
over complex curves, whether or not there is an underlying matrix
model. See also \cite{bkmp} for recent progress in 
this direction.} In this situation, the Calabi-Yau geometry reduces to 
the spectral curve of the matrix model, and the resulting
simplification in the Hodge structure should be the origin of the remaining
discrepancies. Via open-closed duality for matrix models, we 
also anticipate a connection to the larger framework of 
\cite{adkmv}, even though concrete attempts in this direction 
have so far been largely unsuccessful. 

As another example, we mention that the wavefunction properties
of the open topological string amplitudes on local Calabi-Yau
manifolds have been previously discussed in \cite{adkmv,openbps}, 
see also \cite{kashani}. In that context, the main discrepancies
from our work seem to originate from the generic decoupling of open
string moduli, or else the inaccessibility of certain D-brane charge 
sectors, on compact Calabi-Yau manifolds.

We hope that the clarification of the relation of our results with 
those and other works will help to isolate the central structures,
and possibly even shed some light on the much more important problem
of background independence in physical string theory, maybe along 
the lines of \cite{wittenosft,samson1,samson2}.

\subsection{Results}

In \cite{coy}, it was noted that the general solution of the 
extended holomorphic anomaly equation of \cite{extended} can be mapped to 
a solution of the ordinary (BCOV) holomorphic anomaly equation by a 
certain shift of the closed string variables. This observation was
then used to give a proof of the Feynman rule computation of the 
open topological string amplitudes given in \cite{extended}, following
\cite{bcov2}. Other recent work on the holomorphic anomaly for open
strings includes \cite{bonelli,laenge,konishi}, and see \cite{agnt2}
for some earlier work.

Could such a simple relation between the open and closed topological string 
also hold for the {\it physical solutions} of the anomaly equations, 
namely, for the topological amplitudes themselves?
The holomorphic anomaly equations do not constrain the holomorphic dependence
of the topological amplitudes, so there is clearly something non-trivial 
to check. On the other hand, the holomorphic part of the amplitudes is to 
a large extent constrained by a second set of equations, which express 
the statement that adding closed string insertions 
in worldsheet diagrams is equivalent to taking holomorphic (covariant) 
derivatives with respect to the moduli. In this way, one is actually 
reduced to the problem of determining, at each order in perturbation theory, 
a finite number of constants specifying the holomorphic part of the vacuum 
amplitudes. Fixing this ``holomorphic ambiguity'' normally requires 
additional information not contained in the holomorphic anomaly equation 
itself.

It turns out that when the open string amplitudes for non-trivial D-branes 
are shifted as in \cite{coy}, they in fact {\it do not} obey the second,
holomorphic, set of constraints. We hasten to emphasize that this does not
affect the proof of the Feynman rules given in \cite{coy}, because that 
proof only depends on the validity of the antiholomorphic constraints, namely 
the holomorphic anomaly equations. We can further mitigate the 
disappointment by revealing that there is a {\it different shift}
of the open string amplitudes that leads to a solution of both the
holomorphic and the antiholomorphic constraints.

We can give a brief summary of this result in formulas. In the simplest form, the 
BCOV holomorphic anomaly equations \cite{bcov2} are equivalent to a standard 
heat equation \cite{gnp}:
\begin{equation}
\eqlabel{first}
\begin{split}
\biggl[\frac{\del}{\del X^I} - \frac{1}{2} C_{IJK} 
\frac{\del^2}{\del y_J\del y_K}\biggr] &\Psi_{\rm closed} = 0\,, \\
\frac{\del}{\del\bar X^I} \Psi_{\rm closed}  =& 0\,,
\end{split}
\end{equation}
where $\Psi_{\rm closed} := \Psi_{\rm closed}(X^I,y_I)$, as a function 
of the $y_I$, is the generating function of the (closed) topological string 
amplitudes $\Fc g_{i_1, \ldots,i_n} :=  \F g0_{i_1, \ldots, i_n}$, themselves 
functions of homogeneous 
coordinates $X^I$ on the closed string moduli space. In \eqref{first}, 
$C_{IJK}$ is the three-point function on the sphere (Yukawa
coupling), which is the basic data of the closed topological string at
tree level.

As we will show, the extended holomorphic anomaly equations of \cite{extended}
are equivalent to a heat equation extended by a ``convection term,''
\begin{equation}
\eqlabel{exfirst}
\begin{split}
\biggl[\frac{\del}{\del X^I} - \frac{1}{2} C_{IJK} &
\frac{\del^2}{\del y_J\del y_K} - \ii \mu \nu_{IJ} \frac{\del}{\del y_J} \biggr] 
\Psi_{\rm open} = 0\,, \\
&\frac{\del}{\del\bar X^I} \Psi_{\rm open} =0\,,
\end{split}
\end{equation}
for the generating function $\Psi_{\rm open}$ of the topological amplitudes
$\F gh_{i_1, \ldots, i_n}$. Here $\nu_{IJ}$ is (part of) the disk amplitude with 
two bulk insertions, which
is the basic holomorphic data specifying the D-brane background \cite{extended}.
The tensor $\nu_{IJ}$ can be integrated to the superpotential or domain wall
tension, $\nu_{IJ}=\del_I\nu_J=\del_I\del_J \calt$. We also 
introduced a formal real parameter $\mu$
counting the number of worldsheet boundaries.

It is fairly obvious that \eqref{exfirst} can be transformed into \eqref{first}
by a simple shift of the closed string variables, $y_I\to y_I - \ii \mu \nu_I$, 
as anticipated in \cite{coy}.\footnote{More precisely, the shift proposed 
in \cite{coy} would read in the present notation as $y_I\to y_I - \ii \mu (\nu_I - \bar\nu_I)$.
We will discuss the difference between the two shifts 
extensively in section \ref{shifts}.} In other words, given an open 
string background (specified by $\nu_{IJ}$), we define a {\it shifted} 
open string partition function by $\Psi^\nu(X^I,y_I)=\Psi_{\rm open}(X^I,y_I-\ii \mu 
\nu_I)$ (in this notation, $\Psi_{\rm closed}=\Psi^0$). This $\Psi^\nu$ 
then satisfies the ordinary heat equation \eqref{first}, independent of 
$\nu_{IJ}$.

It is now meaningful to ask whether the shifted open string partition function is 
equal to the closed string partition function, in other words, whether $\Psi^\nu
\overset{?}{=}\Psi^0$ is in fact independent of $\nu$. As it turns out,
the answer is in the negative,\footnote{Interestingly, the discrepancy between
$\Psi^\nu$ and $\Psi^0$ arises {\it before} taking into account the 
holomorphic ambiguity.} but it shines in a positive light when viewed 
instead as an answer to a long-standing question raised by Witten's
interpretation of the holomorphic anomaly equation in the context of 
background independence, which we now recall.

\subsection{A positive attitude}

For fixed background $X^I$, Witten proposed \cite{wittenwf} to view the 
topological string partition function $\Psi(X^I,y_I)$, as a function of the
$y_I$, as a ``wavefunction'' specifying a quantum mechanical state in a 
particular presentation of a certain
Hilbert space. Witten's Hilbert space, which we denote by
$\calh_W$, arises from the quantization of the symplectic
vector space of topological ground states of the underlying worldsheet
theory. The choice of background is equivalent to specifying a 
complex polarization of this vector space, and hence a particular
presentation of the wavefunctions. The wavefunction depends on the background,
but in a way that is completely fixed, according to the heat equation \eqref{first},
by the variation of the polarization. The abstract quantum mechanical 
state itself, $|\Psi\rangle\in\calh_W$, is independent of the background.

Our punchline might be clear already. A priori, the physical 
significance of Witten's auxiliary Hilbert space $\calh_W$
is obscure if the closed topological string specifies only one
particular state in it. But if for any D-brane configuration whose 
topological amplitudes satisfy the extended holomorphic anomaly
equation, the shifted partition function $\Psi^\nu$ satisfies the
ordinary heat equation, this means precisely that any such D-brane
specifies a state $|\Psi^\nu\rangle\in\calh_W$. In fact, in section 
\ref{speculations} we will describe evidence that, as $\nu$ varies 
over all possible D-brane configurations, the set of $|\Psi^\nu\rangle$ 
furnishes a basis of $\calh_W$, thus filling Witten's entire Hilbert space 
with life.

It remains to be understood what physical principle selects the basis
of states $|\Psi^\nu\rangle$, and in particular the closed string ground
state $|\Psi^0\rangle$. To give some hints at the nature of this question,
we note that when our topological string is the B-model on some
Calabi-Yau threefold $Y$, then the basis $|\Psi^\nu\rangle$ is,
at least partially, indexed by the set of all possible holomorphic 
curves in $Y$ (see section \ref{speculations} for details).
In other words, understanding the quantum Hilbert space of the
topological B-model on $Y$ involves knowledge of all holomorphic
curves in $Y$. Of course, the topological string knows a great deal
about holomorphic curves on Calabi-Yau threefolds. Remarkably
though, this knowledge arises from studying the A-model on $Y$,
whereas we would here be trying to answer a B-model question
(on the same manifold, not its mirror). This is suggestive of 
an intimate relation between topological A- and B-model on the
same Calabi-Yau manifold, once D-branes are appropriately taken
into account. This basic point has been emphasized by many people, 
beginning with \cite{oooy}, and is an ingredient in the topological
S-duality proposal of \cite{sduality1,sduality2}. We also note that 
speculations along the above lines first appeared in the work of Donagi 
and Markman \cite{doma}.

Let us now close this introduction and start with the derivation of 
the above-mentioned results. We will return to their 
interpretation in section \ref{speculations}, where we'll also
include some more concrete speculations on the relation to BPS 
state counting.

\section{Derivations}

We are interested in the topological string obtained by twisting an
$\caln=2$ superconformal field theory of central charge $\hat c=3$
and all-integral $U(1)$ charges. As shown in \cite{ceva,bcov2} in great
generality, the space of chiral deformations of such a superconformal 
field theory carries the structure of a special K\"ahler manifold,
which we will denote by $\calm$, and which forms the basic holomorphic
arena for the topological string. We will use local coordinates 
$t^i$, $i=1,2,\ldots, n={\rm dim}_\complex(\calm)$.

We note that our conventions in this section differ for convenience from those
in \cite{extended,gnp}, which will result in various factors of $\ii$
appearing differently.

\subsection{Special geometry with D-branes}
\label{specialgeometry}

The central data of special geometry of $\calm$ are the Hodge line bundle 
$\call$, with a Hermitian metric whose curvature is the special K\"ahler 
form of $\calm$, and the Yukawa cubic 
$C$, which is a holomorphic section of $\call^{-2}\otimes 
{\rm Sym}^3 T^*\calm$. The metric on $\call$ is denoted as $e^{-K}$ w.r.t.\ 
some local trivialization, providing a K\"ahler potential for the special 
K\"ahler metric on $\calm$, $G_{i\jb}=\del_i\del_\jb K$.
We will write $D$ generically for the metric-compatible connections
on products of powers of $\call$ and $T$.

A crucial object is the ``bundle of ground states'' which we define as
\begin{equation} \label{gs-bundle}
V_\complex := \call \, \oplus \, \call \otimes T\calm \, \oplus \, 
\bar\call \otimes \bar T\calm \, \oplus \, \bar\call\,,
\end{equation}
where $T\calm$ is the holomorphic tangent bundle of $\calm$. $V_\complex$ 
has an obvious conjugation operator $\bar{\cdot}$ which defines a real 
sub-bundle $V_\reals$, a Hermitian metric $\IP{\cdot|\cdot}$ induced 
from those on $T\calm$ and $\call$, and an antisymmetric bilinear form 
$\IP{\cdot,\cdot}$ defined by
\begin{equation}
\IP{a,b} := \IP{\bar{a} | \sigma b} \quad \text{where} \quad \sigma := \begin{cases} 
+i \text{ on } \call \, \oplus \, \bar\call \otimes \bar T\calm\,, \\
-i \text{ on } \bar\call \, \oplus \, \call \otimes T\calm\,. \\
\end{cases}
\end{equation}
We introduce an operator $\hC: T\calm \otimes V_\complex \to V_\complex$;
for $v \in T\calm$, $\hC(v) \in \End(V_\complex)$ maps each space in \eqref{gs-bundle} to its successor,
\begin{equation}
\hC(v) |_\call = \cdot \otimes v\,, \quad
\hC(v) |_{\call \otimes T\calm} = C(v) e^K G^{-1}\,, \quad
\hC(v) |_{\bar\call \otimes \bar T\calm} = G(v,\cdot)\,, \quad
\hC(v) |_{\bar\call} = 0\,.
\end{equation}
Similarly $\bar{\hC}(v)$ maps each space in \eqref{gs-bundle} to its predecessor.
Using $\hC$ we can define the ``Gauss-Manin connection'' on $V_\complex$ in terms 
of its holomorphic and antiholomorphic parts,
\begin{equation}
\eqlabel{gm}
\nabla = D - \ii \hC\,, \quad \bar\nabla = \bar{D} + \ii \bar{\hC}\,.
\end{equation}
The connection preserves $V_\reals$.
Moreover, it is flat, as one verifies using the special geometry formula for the curvature of $G$.
Hence it makes $V_\complex$ into a holomorphic vector bundle, and
there is a natural filtration by holomorphic subbundles
\begin{equation}
0 \subset F^3 V_\complex \subset F^2 V_\complex \subset F^1 V_\complex \subset F^0 V_\complex = V_\complex\,,
\end{equation}
where $F^k V_\complex$ is the sum\footnote{This is a direct sum of complex vector bundles; we emphasize however 
that its holomorphic structure induced from the Gauss-Manin connection is not a direct sum, because 
$\bar{\hat C}$ mixes the summands.}
of the first $4-k$ summands in \eqref{gs-bundle}.

When the special K\"ahler manifold arises from twisting of an $\caln=2$
field theory, we have more data than what was mentioned above. The first extra datum
is a twisted chiral ring, which in particular allows us to 
define the Euler characteristic,
\begin{equation}
\eqlabel{eulerchar}
\chi := 2n-2{\rm dim}\bigl(\text{$(a,c)$-ring}\bigr)\,.
\end{equation}
The second extra datum is a lattice $V_\zet^* \subset V^*_\reals$ preserved by the Gauss-Manin connection.
We assume that $V_\zet^*$ is self-dual
for the skew-symmetric pairing.  Then we can choose an integral basis 
$\{\alpha^I, \beta_I\}$ for $V_\zet^*$, obeying 
\begin{equation}
\IP{\alpha^I, \alpha^J} = 0\,, \quad \IP{\alpha^I, \beta_J} = \delta^I_J\,, \quad \IP{\beta_I, \beta_J} = 0\,.
\end{equation}
Having fixed such a basis and a local section $\Omega$ of $\call$ one gets $2n+2$ functions 
on $\calm$, the ``periods''
\begin{equation}
X^I = \alpha^I(\Omega)\,, \quad F_I = \beta_I(\Omega)\,,
\end{equation} 
with $I = 0, \dots, n$.
The K\"ahler potential on $\calm$ can then be written as
\begin{equation}
\ee^{-K} = \ii \bigl(X^I \bar F_I - \bar X^I F_I\bigr)\,.
\end{equation}

Now we consider the open string sector.  As argued in \cite{mowa,extended}, 
certain data of topological D-brane configurations on Calabi-Yau threefolds 
can be encoded holomorphically in the Hodge theoretic concept of {\it normal
functions}. By definition, a normal function $\nu$ is a
holomorphic section of the intermediate Jacobian fibration
\begin{equation}
\eqlabel{jacfibration}
J^3 :=  V_\complex / (V_\zet + F^2 V_\complex) \simeq (F^2 V_\complex)^* / V_\zet^*
\end{equation}
satisfying Griffiths transversality,
\begin{equation}
\nabla \tilde \nu \in F^1 V_\complex\,,
\end{equation}
where $\tilde\nu$ is any lift of $\nu$ to $V_\complex$, as well as certain
growth conditions at infinities in $\calm$ that will not be of central
importance for our considerations. We shall not review here the discussion
of the relation of D-branes with normal functions. We emphasize, 
however, that one of the fundamental ingredients in this identification is 
the decoupling of any continuous open string moduli from the topological string 
amplitudes. This might not apply in the most general situation, but we shall
assume it in what follows. We also point out that the normal function need 
not capture the full information contained in any given D-brane. To simplify 
our presentation, we use $\nu$ as a shorthand for the D-brane configuration.

The central point of \cite{extended} was the identification of the
open string counterpart of the Yukawa cubic, which as reviewed above
is the defining holomorphic data of the closed topological string
at tree level. Recall that the Yukawa coupling is computed as the
three-point function on the sphere. The open analogue is the
two-point function on the disk, given geometrically by a particular non-holomorphic 
lift of the Griffiths infinitesimal invariant of the normal function 
$\nu$. A pedagogical reference on normal functions and their infinitesimal
invariants is \cite{cime}. This lift is a section of $\call^{-1}\otimes 
{\rm Sym}^2 T^*\calm$, and can be written as%
\footnote{This expression differs by a factor of $\ii$ from the
corresponding expression in \cite{extended}. This factor can be traced back
to a different convention for the Gauss-Manin connection, see
\eqref{gm}.}
\begin{equation} \label{delta-abs}
\Delta_{ij} = \IP{\Omega, \nabla_i \nabla_j \tilde\nu} = D_i D_j \calt + 
\ii C_{ijk} e^K G^{k\kb} D_\kb \bar \calt\,,
\end{equation}
where $\calt$ is a section of $\call^{-1}$ given by
\begin{equation}
\calt = \IP{\Omega, \nu}
\end{equation}
and we have chosen the unique \textit{real} lift $\tilde\nu \in V_\reals$.

Crucially, $\Delta_{ij}$ is not a holomorphic section.
Instead, it satisfies a holomorphic anomaly equation,
\begin{equation}
\eqlabel{diskhan}
\del_\kb \Delta_{ij} = \ii C_{ijk} \bar \Delta_\kb^k\,,
\end{equation}
where indices are raised with the metric, $\bar\Delta_{\kb}^k
= e^K G^{k \jb} \bar\Delta_{\kb\jb}$.

A fundamental example of the structure described above is provided by Type II string
theory on a Calabi-Yau threefold $Y$. In that case: $\calm$ is the moduli space 
of complex structures on $Y$; $V_\complex = H^3(Y,\complex)$ (similarly for
$V_\reals$, $V_\zet$); the decomposition \eqref{gs-bundle} is the Hodge
decomposition; and the Gauss-Manin connection is the standard flat structure
provided by the deformation invariance of integer homology. The basic example
of a normal function comes from a pair of homologically equivalent holomorphic 
curves $C_+$, $C_-$ in $Y$ varying over $\calm$. Over every point in
$\calm$ we pick a three-cycle $\Gamma$ in $Y$ such that $\del\Gamma=C_+-C_-$.
We obtain an element $\nu$ of $(F^2V_\complex)^* \simeq (H^{3,0}\oplus H^{2,1})^*$
by associating to any $(3,0)$ or $(2,1)$-form $\omega$ the chain integral
\begin{equation}
\eqlabel{abeljacobi}
\langle\omega,\nu\rangle = \int_\Gamma\omega
\end{equation}
viewed as a function over moduli space. The chain integral of cohomology classes 
in \eqref{abeljacobi} is well-defined by Dolbeault's theorem and the 
holomorphicity of $\del\Gamma$. Changing  the choice of $\Gamma$ by a closed 
three-cycle, the integral changes by a period, which precisely accounts for 
the quotient by $V_\zet^*$ in \eqref{jacfibration}.

The association \eqref{abeljacobi} is known as the Abel-Jacobi map. 
It was first used in the context of mirror symmetry with 
D-branes on non-compact Calabi-Yau manifolds in \cite{agva,akv}, and in a 
compact setting in \cite{open,mowa}. The relation \eqref{diskhan} can be
viewed as an open string analogue of special geometry \cite{extended}.
See \cite{lmw1,lmw2} for a concurrent proposal.

\subsection{Holomorphic anomaly of topological string amplitudes}

We now fix some topological string background with D-branes
characterized at tree-level by a normal function $\nu$ living
over the special K\"ahler moduli space $\calm$. We are interested
in the perturbative topological string amplitudes $\F gh_{i_1,i_2,\ldots,i_n}$. 
These amplitudes are defined by integrating over the moduli space of Riemann 
surfaces of genus $g$, with $h$ boundary components, the appropriate correlators 
of the underlying worldsheet theory. The indices $i_1,i_2,\ldots, i_n$ 
stand for closed string insertions. See, \eg, \cite{extended} for
details. 

As mentioned in the introduction, the $\F gh_{i_1,i_2,\ldots,i_n}$ satisfy 
two sets of iterative relations. 
The more obvious relations, which are holomorphic in nature, iteratively relate
the amplitudes with insertions to those without. The amplitudes are
non-zero for $6g+3h+2n-6\ge 0$, and we have in this case
\begin{equation}
\eqlabel{firstset}
\F gh_{i_1,i_2,\ldots,i_{n+1}} = D_{i_{n+1}} \F gh_{i_1,\ldots,i_{n}}\,.
\end{equation}
The amplitudes with $6g+3h+2n-6<0$ (all of which are tree-level
amplitudes) are customarily set to zero, and we note that the vacuum 
amplitudes $\F gh$ for $2g+h-2\ge 0$, the 
sphere three-point function $\F 00_{ijk}=C_{ijk}$, and the disk two-point 
function, $\F 01_{ij}=\Delta_{ij}$ are not constrained by \eqref{firstset}
in any way.

The second set of relations are less obvious, but equally fundamental, 
and relate amplitudes on different worldsheet topologies,
\begin{multline}
\eqlabel{secondset}
\del_\ib \F gh_{i_1,\ldots,i_n} = 
\frac 12 \sum_{\topa{g_1+g_2=g}{h_1+h_2=h}} \bar{C}_{\ib}^{jk}
\sum_{s,\sigma \in S_n} \frac{1}{s!(n-s)!}
\F {g_1}{h_1}_{j,i_{\sigma(1)},\ldots,i_{\sigma(s)}} 
\F {g_2}{h_2}_{k,i_{\sigma(s+1)},\ldots,i_{\sigma(n)}} + \\
\frac 12 \bar{C}_{\ib}^{jk} \F {g-1}h_{j,k,i_1,\ldots,i_n} 
+ i \bar{\Delta}_{\ib}^j \F g{h-1}_{j,i_1,\ldots,i_n}
-(2g+h-2+n-1)\sum_{s=1}^n G_{i_s\ib}
\F gh_{i_1,\ldots,i_{s-1},i_{s+1},\ldots,i_n}\,.
\end{multline}
These equations are valid for all $2g+h+n-2> 0$, except for the one-point 
functions at one-loop $(g,h,n)=(1,0,1)$ or $(0,2,1)$, for which we have 
an additional term on the right hand side:
\begin{equation}
\eqlabel{oneloop}
\begin{split}
\del_\ib \F 10_j &= \frac 12 C_{jkl} \bar C_{\ib}^{kl} - \Bigl(\frac\chi{24}-1
\Bigr) G_{j\ib}\,, \\
\del_\ib \F 02_j &= i \Delta_{jk} \bar\Delta_{\ib}^k - \frac N2
G_{j\ib}\,,
\end{split}
\end{equation}
where $\chi$ is the Euler characteristic \eqref{eulerchar}, and $N$ is 
the number of open string Ramond ground states of zero charge. 
The one-loop vacuum amplitudes are not constrained by \eqref{secondset}
directly, but indirectly by \eqref{firstset} and \eqref{oneloop}.
The holomorphicity of the sphere three-point function $\del_\lb C_{ijk}=0$, 
as well as the holomorphic anomaly of the disk two-point function, 
\eqref{diskhan}, appear as special cases of \eqref{secondset}. (Here, one
has to use the vanishing of the tree-level amplitudes with few insertions.)

The two sets of equations \eqref{firstset} and \eqref{secondset} together
with their exceptional modification at one-loop and tree-level can be
summarized more concisely in two ``master equations'' by introducing a 
certain generating function,
\begin{equation}
\eqlabel{certain}
\Psi(t^i,\bar t^\ib;x^i,\lambda^{-1})
= \lambda^{\frac{\chi}{24}-1- \mu^2 \frac N2}
\exp\biggl[ \sum_{\topa{g,h,n}{2g+h+n-2>0}}
\frac{\lambda^{2g+h+n-2}}{n!} \mu^h \F gh_{i_1,\ldots,i_n} x^{i_1}\cdots x^{i_n}
\biggr]\,.
\end{equation}
With this definition, $\Psi$ is a section of the pullback of
$\call^{\frac{\chi}{24}-1-\mu^2 \frac N2}$ to the total space of $(\call \, \oplus \, \call \otimes T\calm) \to \calm$. 
The variables $x^i$ are coordinates on $\call \otimes T\calm$, the inverse string coupling
$\lambda^{-1}$ is a coordinate on $\call$, and $\mu$ is a real parameter that keeps track of the number of
boundaries in the expansion \eqref{certain}.  $\Psi$ is holomorphic on each fiber but not holomorphic on the total space.

It is not hard to check that the relations \eqref{secondset} are equivalent to 
the following equation satisfied by the generating function $\Psi$:
\begin{equation}
\eqlabel{firsthan}
\biggl[\del_{\ib}-\frac 12 \bar C_{\ib}^{jk}\frac{\del^2}{\del x^j\del x^k}
- G_{j\ib} x^j \frac{\del}{\del\lambda^{-1}} -
\ii \mu \bar\Delta_{\ib}^j \frac{\del}{\del x^j}\biggr]\Psi = 0\,.
\end{equation}
Also, the relations \eqref{firstset}, together with the statements about
the amplitudes with $2g+h+n-2\le 0$ (which are absent from \eqref{certain}), 
are equivalent to
\begin{equation}
\eqlabel{secondhan}
\begin{split}
\biggl[\del_i - \Gamma_{ij}^k x^j \frac{\del}{\del x^k} &+ \del_i K
\Bigl( \lambda^{-1}\frac{\del}{\del\lambda^{-1}} + x^k\frac{\del}{\del x^k}
+\frac \chi{24}-1- \mu^2 \frac N 2 \Bigr) \\
& - \lambda^{-1} \frac{\del}{\del x^i} + \F 10_i + \mu^2 \F 02_i + \frac 12
C_{ijk} x^j x^k + \mu \Delta_{ij} x^j\biggr] \Psi = 0\,.
\end{split}
\end{equation}
In the rest of this section, we will rewrite the equations \eqref{firsthan}
and \eqref{secondhan} in various ways. The purpose is to show that the
underlying structure is fairly simple, especially when viewed from the
point of view of the so-called ``large phase space''. Readers interested 
primarily in the conceptual questions might be able to skip directly to
section \ref{discussion} on page \pageref{discussion}.

\subsection{The large phase space}

To construct the large phase space \cite{dvv}, 
one first replaces the moduli space $\calm$
by the total space $\tcalm$ of the line bundle $\call$ minus the
zero section.  One reason for introducing $\tcalm$ is that it comes equipped with
very convenient coordinates, namely the $X^I$ we considered before.
In other words, a choice of nowhere vanishing section of $\call$ defines an embedding of $\calm$
into $\tcalm$ via
\begin{equation}
\eqlabel{embed}
X^I=X^I(t^i)\,,
\end{equation}
with inverse projection
\begin{equation}
\eqlabel{inverse}
t^i = t^i(X^I)
\end{equation}
which is homogeneous of degree $0$,
\begin{equation}
X^I \frac{\del}{\del X^I} t^i(X^I) = 0\,.
\end{equation}
All quantities we consider are homogeneous of fixed degree under the overall rescaling of the $X^I$.

As complex vector bundles over $\calm$,
$\call \, \oplus \, \call \otimes T\calm \simeq F^2 V_\complex$; the pullback of $F^2 V_\complex$ to $\tcalm$ is isomorphic to $T \tcalm$.  The map from the pullback of $\call \, \oplus \, \call \otimes T\calm$
to $T \tcalm$ is\footnote{For convenience, this formula differs from that of \cite{gnp} by an eighth root of unity.}
\begin{equation} \label{coc}
z^I = 2\,(\lambda^{-1} X^I + x^i X^I_{;i} )
\end{equation}
where $z^I$ is the coordinate on the fiber of $T \tcalm$, and we defined
\begin{equation} \eqlabel{we-defined}
X^I_i := \del_i X^I\,, \qquad X^I_{;i} := X^I_i + \del_i K X^I\,.
\end{equation}
Altogether, we have changed from the ``small phase space'' coordinates 
$(t^i, x^i, \lambda^{-1})$ on $(\call \, \oplus \, \call \otimes T\calm)
\to \calm$ to the ``large phase space'' coordinates $(X^I, z^I)$ on $T \tcalm$.

Next we describe special geometry from the large phase space point of view,
beginning with the closed string.
The basic holomorphic data is encoded in the prepotential
\begin{equation}
F := \frac 12 X^I F_I
\end{equation}
and its first few derivatives,
\begin{equation}
F_I = \del_I F\,, \qquad \tau_{IJ} := \del_I\del_J F\,, \qquad
C_{IJK} := \del_I\del_J\del_K F\,.
\end{equation}
The homogeneity of $F$ implies a useful relation for the Yukawa coupling,
\begin{equation}
\eqlabel{homoyuk}
X^I C_{IJK} = 0\,.
\end{equation}

Now we describe the large phase space version of the open string data, namely, normal functions
and their infinitesimal invariants. The domain wall 
tension $\calt = \langle \Omega, \nu \rangle$ 
is a period-like object, homogeneous of degree $1$ in the large phase space:
\begin{equation}
\calt = X^I\nu_I\,, \qquad {\rm where}\; \nu_I:=\del_I\calt\,.
\end{equation}
The large phase space infinitesimal invariant is
\begin{equation} \eqlabel{delta-big}
\Delta_{IJ} := \IP{\Omega,\del_I \del_J \nu} = \nu_{IJ} - C_{IJ}^K \Im\nu_K\,,
\end{equation}
where of course
\begin{equation}
\nu_{IJ} := \del_I\del_J\calt\,,\qquad {\rm with}\; X^I \nu_{IJ} = X^I\Delta_{IJ}=0\,.
\end{equation}
The last relation is very similar to \eqref{homoyuk} satisfied by the Yukawa 
coupling.

In Appendix \ref{applargesmall} we give some useful relations between the large and 
small phase space data.

\subsection{The anomaly equations in the large phase space}

In this section we will rewrite the anomaly equations in the large phase space,
which turns out to be the most convenient setting for understanding the 
relation between the open and closed anomaly equations.

We follow \cite{verlinde} and first solve the holomorphic anomaly equations
for the one-loop amplitudes \eqref{oneloop}. Recall that the torus anomaly 
can be integrated to \cite{bcov1}
\begin{equation}
\eqlabel{torusint}
\F 10 = -\frac 12 \log\det\Im\tau_{IJ} - \Bigl(\frac\chi{24}-1\Bigr) K + 
f^{(1,0)} + \bar f^{(1,0)}\,,
\end{equation}
where $f^{(1,0)}$ is a holomorphic ambiguity. The holomorphic anomaly equation
of the annulus (as well as higher $(g,h)$) can be integrated by a procedure 
very similar to that in \cite{bcov2}, see \cite{extended}. To this end, let 
us introduce the large phase space analogue, $\delta^J$, of the ``terminator'' of 
\cite{extended}. This quantity is defined by the equation
\begin{equation}
\eqlabel{property}
\bar\Delta_I^J = \bar\del_I \delta^J\,,
\end{equation}
which can be locally solved by
\begin{equation}
\eqlabel{locally}
\delta^J = -2\ii \Im\tau^{JK} \Im\nu_K = \Im\tau^{JK} (\bar\nu_K - \nu_K)\,.
\end{equation}
This choice of $\delta^J$ also satisfies
\begin{equation}
\del_I \delta^J = - \Delta_I^J.
\end{equation}
For future reference, we also record the holomorphic anomaly of the 
disk two-point function in large phase space:
\begin{equation}
\eqlabel{future}
\bar\del_K \Delta_{IJ} = -\frac\ii 2 C_{IJL}
\bar\del_K \delta^L = -\frac \ii 2 C_{IJL} \bar\Delta_K^L\,.
\end{equation}
With these definitions, the holomorphic anomaly equation of the
annulus, which in the large phase space takes the form
\begin{equation}
\bar\del_J\del_I \F 02 = - \frac{\ii}{2} \Delta_{IK} \bar\Delta_J^K + \frac N2
\bar\del_J\del_I K\,,
\end{equation}
can be integrated to 
\begin{equation}
\eqlabel{canbe}
\begin{split}
\F 02_I &= - \frac{\ii}{2} \Delta_{IK}\delta^K + \frac 1 8 C_{IKL}
\delta^K\delta^L + \frac N2 \del_I K + \del_I f^{(0,2)}\,, \\
\F 02 &= \frac i4 \delta^K \Im\tau_{KL} \delta^L + \frac N2 K + f^{(0,2)}+\bar f^{(0,2)}\,,
\end{split}
\end{equation}
where $f^{(0,2)}$ is another holomorphic ambiguity.

We absorb $f^{(1,0)}$ and $f^{(0,2)}$ into a
redefinition of the generating function $\Psi$:
\begin{equation}
\Psi \to \ee^{-f^{(1,0)} - \mu^2 f^{(0,2)}} \Psi\,.
\end{equation}
This should be interpreted as $\Psi_{\rm old} = \ee^{-f^{(1,0)} - \mu^2 f^{(0,2)}} \Psi_{\rm new}$;
we will use this notation repeatedly in the next few sections.

It is then straightforward, using the formulas given in Appendix \ref{applargesmall}, 
to show that the equations \eqref{firsthan}
and \eqref{secondhan} become respectively
\begin{equation}
\eqlabel{become}
\begin{split}
\Bigl[\delbar_{I}  -  \frac 1 2 \bar C_I^{JK} \frac{\del^2}{\del z^J\del z^K}
 + & \ii \mu \bar \Delta_I^J \frac{\del}{\del z^J}\Bigr] \Psi = 0\,, \\
\Bigl[\del_I - \frac 12 \del_I \log\det\Im\tau_{IJ} + 
\frac \ii 2C_{IJ}^K z^J \frac{\del}{\del z^K} &+\frac 1 8 C_{IJK} z^J z^K \\
+ \frac 12 \mu \Delta_{IJ} z^J &
- \frac \ii 2 \mu^2 \Delta_{IJ}\delta^J  + \frac 1 8 \mu^2 
C_{IJK}\delta^J\delta^K\Bigr]\Psi = 0\,.
\end{split}
\end{equation}
The closed string version of these equations was first obtained in \cite{dvv}.
Now note that by using
\begin{equation}
\eqlabel{using}
\Delta_{IJ} = \nu_{IJ} - C_{IJ}^K \Im\nu_K = \nu_{IJ} - \frac\ii 2 C_{IJK} \delta^K
\end{equation}
we can rewrite the last four terms in \eqref{become} as
\begin{equation}
\begin{split}
\frac 1 8 C_{IJK} z^J z^K + \frac 1 2 \mu &\nu_{IJ} z^J -\frac \ii 4  \mu C_{IJK} z^J \delta^K
-\frac \ii 2 \mu^2 \nu_{IJ}\delta^J - \frac 1 8 \mu^2 C_{IJK} \delta^J\delta^K \\
&= \frac 1 8 C_{IJK} (z^J - \ii \mu \delta^J)(z^K-\ii \mu \delta^K)
+\frac 1 2 \mu \nu_{IJ} (z^J - \ii \mu \delta^J)\,.
\end{split}
\end{equation}
So if we now resubstitute
\begin{equation}
\eqlabel{good}
\Psi \to \sqrt{\det\Im\tau_{IJ}} \exp\Bigl[\frac \ii 4(z^J-\ii \mu\delta^J)\Im\tau_{JK}
(z^K-\ii \mu \delta^K)\Bigr] \Psi
\end{equation}
and use the above relations for $\Delta_{IJ}$, $\delta^J$, etc., the
equations take the form
\begin{equation}
\begin{split}
\Bigl[ \bar\del_I - \frac\ii 2 \bar C_{IJ}^K z^J \frac{\del}{\del z^K}
- \frac 1 2 \bar C_I^{JK} \frac{\del^2}{\del z^J\del z^K}
+ &\ii \mu\bar\nu_I^J \frac{\del}{\del z^J}\Bigr] \Psi =0\,, \\
\Bigl[\del_I + \frac \ii 2 C_{IJ}^K z^J \frac{\del}{\del z^K} \Bigr] \Psi &=0\,.
\end{split}
\end{equation}
Finally, we change variables to $\bar{y}_I = \Im\tau_{IJ} z^J$ \cite{gnp}.
Geometrically this amounts to considering $\Psi$ as defined on $T^*\tcalm$
instead of $T \tcalm$.  We then arrive at the simplest form of the anomaly 
equations,
\begin{equation}
\eqlabel{arrive}
\begin{split}
\Bigl[ \bar\del_I - \frac 1 2 \bar C_{IJK} \frac{\del^2}{\del \bar{y}_J\del \bar{y}_K}
&+ \ii \mu \bar\nu_{IJ} \frac{\del}{\del \bar{y}_J} \Bigr] \Psi = 0\,, \\
 \del_I \Psi &=0\,.
\end{split}
\end{equation}
For completeness, we summarize the sequence of redefinitions of $\Psi$
leading from \eqref{certain} to \eqref{arrive}:
\begin{equation}
\eqlabel{redefine}
\Psi_{\eqref{arrive}} = \frac{1}{\sqrt{\det\Im\tau_{IJ}}} 
\exp\Bigl[-\frac \ii 4 (z^J - \ii \mu\delta^J)\Im\tau_{JK} (z^K-\ii \mu\delta^K)+
f^{(1,0)} + \mu^2 f^{(0,2)}\Bigr] \Psi_{\eqref{certain}}\,.
\end{equation}

After all these transformations $\Psi$ has turned out to be purely antiholomorphic.
We can take $\Psi \to \overline{\Psi}$ to get 
\begin{equation}
\eqlabel{arrive-hol}
\begin{split}
\Bigl[ \del_I - \frac 1 2 C_{IJK} \frac{\del^2}{\del y_J\del y_K}
&- \ii \mu \nu_{IJ} \frac{\del}{\del y_J} \Bigr] \Psi = 0\,, \\
 \bar \del_{I} \Psi &=0\,.
\end{split}
\end{equation}

To conclude this section we briefly discuss the global properties of $\Psi$.  Before the redefinition 
$\Psi$ represented a section of the pullback of $\call^{\frac{\chi}{24}-1-\mu^2 \frac N2}$, \ie\ 
under a change of local section $\Omega \to e^f \Omega$ for $\call$ it transformed by
$\Psi \to e^{(\frac{\chi}{24}-1-\mu^2 \frac N2)f} \Psi$.  After the redefinition this 
transformation is canceled by the explicit transformations of $f^{(1,0)}$ and $f^{(0,2)}$
determined by \eqref{torusint} and \eqref{canbe}.  However, $f^{(1,0)}$, $f^{(0,2)}$,
$z^J$, $\delta^J$ and $\tau$ appearing in \eqref{redefine} are all defined using a symplectic basis 
for $V_\zet$, and such a choice cannot be made globally on $\tcalm$ because $V_\zet$ can have 
global $Sp(2n+2,\zet)$-valued monodromies; so the new $\Psi$ has to be considered as a section 
of a bundle which transforms appropriately under $Sp(2n+2,\zet)$, \ie\ as a modular form.

\section{Discussion}
\label{discussion}

The rewriting of the holomorphic anomaly equations in
the last section makes the redefinition of variables
which removes the open string data from the equation completely
transparent. Shifting\footnote{Again, this shift should be interpreted
as $(\bar y_I)_{\rm old}=(\bar y_I)_{\rm new} +\ii\mu\bar\nu_I$.}
\begin{equation}
\eqlabel{bigshift}
\bar y_I \to \bar y_I +\ii \mu \bar \nu_I
\end{equation}
in \eqref{arrive} eliminates $\nu$ from both equations, and maps them
onto the ordinary heat equations satisfied by the closed topological
string amplitudes \cite{gnp}. 

The existence of a shift removing open string data was first noted in 
\cite{coy}. With slightly different conventions, it was shown there that
a general solution of the master holomorphic anomaly equation 
\eqref{firsthan} can be mapped to a solution of the master equation 
with $\Delta_{ij} :=  0$ (which is the master equation of BCOV) by 
shifting the closed string variables. The fate of the second equation 
\eqref{secondhan} was however not analyzed in \cite{coy}. Instead, it 
was shown, using the above observation and the techniques of \cite{bcov2}, 
that the solution of the perturbative holomorphic anomaly equations 
\eqref{secondset} can be written in a diagrammatic fashion 
using Feynman rules as noticed in \cite{extended}.

In this section, we will first discuss the difference between the shift 
\eqref{bigshift} and the shift of \cite{coy}, at the level of the
equation and at the level of the Feynman rules. We will then answer the 
question whether the simple shift maps open to closed topological string 
also at the level of the topological string amplitudes, which are the
physically relevant solutions of the anomaly equations.

\subsection{Shifts}
\label{shifts}

The shift studied in \cite{coy} reads in the small phase space
\begin{equation}
\eqlabel{smallshift}
x^i\to x^i - \ii \mu \bar\Delta^i\,,\qquad\qquad\lambda^{-1} \to \lambda^{-1} 
+ \ii \mu\bar\Delta \,,
\end{equation} 
where $\bar\Delta^i$, $\bar\Delta$ are the terminators of \cite{extended} in the
small phase space, satisfying
\begin{equation}
\eqlabel{full}
\bar\del_\jb \bar\Delta^i= \bar\Delta_\jb^i \,,\qquad
\bar\del_\jb \bar\Delta = G_{\jb i} \bar\Delta^i\,,
\end{equation}
where $\bar\Delta_\jb^i = e^K G^{i\ib}\bar\Delta_{\jb\ib}$. 

To understand the 
relation between \eqref{bigshift} and \eqref{smallshift}, we recall that the 
Griffiths infinitesimal invariant is the sum of two terms, which in small 
phase space are $\Delta_{ij} = \Delta^{(1)}_{ij}
+ \Delta^{(2)}_{ij}$, with $\Delta^{(1)}_{ij}= D_iD_j\calt$ and
$\Delta^{(2)}_{ij} = \ii C_{ij}^{\bar k} D_{\kb}\bar\calt$. In large phase space,
the corresponding two terms are $\Delta_{IJ}=  \Delta^{(1)}_{IJ} + 
\Delta^{(2)}_{IJ}$, with $\Delta^{(1)}_{IJ}= \nu_{IJ} + \frac\ii 2 
C_{IJ}^{K}\nu_K$, $\Delta^{(2)}_{IJ} = -\frac \ii 2 C_{IJ}^{K} \bar\nu_K$. 
From this, we see that our shift \eqref{bigshift} is by a potential for only
the first term, $\bar\del_I \bar\nu^{J} = (\bar\Delta^{(1)})_I^J$,
whereas the shift \eqref{smallshift} is by a potential for the full 
infinitesimal invariant, see \eqref{full}.

In other words, the large phase space analogue of \eqref{smallshift} would
be
\begin{equation}
z^J \to z^J + \ii \mu \delta^J = z^J + \ii \mu (\bar\nu^J-\nu^J)\,.
\end{equation}
It is not hard to see that under this shift, the open string
data is eliminated only from the first equation in \eqref{become},
whereas the second equation remains with an additional term, 
which in the large phase space reads
\begin{equation}
\left( - \ii \mu \nu_I^J \frac{\partial}{\partial z^J} - \frac12 \mu\nu_{IJ} z^J \right) \Psi\,.
\end{equation}
Obviously, one may also write this term in the small phase space. It is
worthwhile pointing out that the difference between the shift \eqref{bigshift} 
and \eqref{smallshift} can not be absorbed into a holomorphic ambiguity inherent 
in the definition of the terminators.

The proof of the Feynman rule expansion of open topological string amplitudes
given in \cite{coy} relies on the same auxiliary finite-dimensional 
quantum system used in \cite{bcov2}. The dynamical variables are the
$x=(x^i,\lambda^{-1})$. The quadratic part of 
the action is $Q(x,x)$, where $Q$ is an inverse to the collection
of propagators $S^{ij}$, $S^{i} :=  S^{i\lambda^{-1}}$, $S :=  
S^{\lambda^{-1}\lambda^{-1}}$. The interactions of the system are 
given by $\log\Psi(t^i,\bar t^\ib;x^i-\ii \mu\bar\Delta^i,\lambda^{-1}+\ii 
\mu\bar\Delta)$ where $\Psi$ is defined in \eqref{certain} and the variables 
are shifted as in \eqref{smallshift}. This includes the vacuum amplitudes 
$\F gh$, and as non-trivial interaction vertices the infinite collection 
of topological amplitudes $\F gh_{i_1,\ldots,i_n}$ with $n\ge 1$.
As pointed out in \cite{coy}, the shift \eqref{smallshift} generates
the possibility of terminating indices on $\bar\Delta^i$, $\bar\Delta =: 
\bar\Delta^{\lambda^{-1}}$, or in other words, it introduces a background
vev for $(x^i, \lambda^{-1})$. This precisely reproduces the Feynman
rules noticed in \cite{extended}.

Our result is that to completely remove the open string data from both
equations, one should shift only by \eqref{bigshift} and absorb the
second term, originating from $(\bar\Delta^{(2)})_I^J$, into the redefinition
of $\Psi$ given in \eqref{redefine}. This has a simple interpretation in 
terms of the finite-dimensional system discussed above. Namely, we give 
a smaller vev to $(x^i, \lambda^{-1})$ and instead
introduce additional vertices $\sim D_i\calt x^i$. The latter also behave 
like a tadpole, so that the overall Feynman rules are unaffected.

Referring to the perturbative holomorphic anomaly equations in \eqref{secondset}, 
we notice that the second part of the infinitesimal invariant enters
as if it were on equal footing with the closed string degenerations. Namely,
it is of the form $(\bar\Delta^{(2)})_{\ib}^j = -i \bar{C}_{\ib}^{jk} D_k \calt$,
where $D_k\calt$ could be viewed as a disk one-point function. The first part, 
$(\bar\Delta^{(1)})_{\ib}^j$, however, can not be treated in this way 
and has to be accounted for in the master equation by a shift of variables 
\eqref{bigshift}.

We remark that the holomorphic anomaly equations for open topological strings
derived in \cite{eynard,marcos1} using matrix model duals to certain local
Calabi-Yau manifolds appear to contain only the first type of contribution
described in the previous paragraph, namely that originating from 
$\bar\Delta^{(2)}_{\bar i \bar j}$. It would be interesting to understand
the Hodge theoretic origin of this simplification, as well as possible 
implications for open/closed duality in this context.

\subsection{Solutions}

We can now answer the question whether the simple shift of variables that 
maps the extended holomorphic anomaly equations to the ordinary BCOV
equations will also transform correctly the actual topological string 
amplitudes. By studying the perturbative expansion in the small phase 
space, we will see that this possibility is in fact not realized.

The small phase space analogue of the shift \eqref{bigshift} is
\begin{equation}
x^i \to x^i - \ii \mu \epsilon^i\,, \qquad \lambda^{-1}\to 
\lambda^{-1} + \ii \mu\epsilon\,,
\eqlabel{sbigshift}
\end{equation}
where $\epsilon^i$, $\epsilon$ are potentials for the first part of the infinitesimal
invariant. They are defined by the equations
\begin{equation}
\begin{split}
\del_\ib \epsilon^j &= \ee^K G^{j\kb} D_{\kb} D_{\ib} \bar \calt \,, \\
\del_\ib \epsilon &= G_{\ib j} \epsilon^j \,,
\end{split}
\end{equation}
which can easily be solved by choosing
\begin{equation}
\begin{split}
\epsilon &= \ee^K \bar \calt \,, \\
\epsilon^i &= G^{i\jb} \ee^K D_{\jb} \bar \calt \,.
\end{split}
\end{equation}
After the shift \eqref{sbigshift}, the small
phase space holomorphic anomaly equations \eqref{firsthan} and
\eqref{secondhan} are transformed into
\begin{equation}
\begin{split}
\biggl[ \del_\ib - \frac 12 \bar C_{\ib}^{jk} \frac{\del^2}{\del x^j\del x^k}
-  G_{j\ib} x^j &\frac{\del}{\del\lambda^{-1}} - \mu \bar C_{\ib}^{jk} 
D_k\calt \frac{\del}{\del x^j}\biggr] \Psi =0 \,, \\
\biggl[\del_i - \Gamma_{ij}^k x^j \frac{\del}{\del x^k} + \del_i K 
\Bigl( \lambda^{-1}& \frac{\del}{\del\lambda^{-1}} + x^k\frac{\del}{\del x^k}
+\frac \chi{24}-1 \Bigr) - \lambda^{-1} \frac{\del}{\del x^i} + \F 10_i + \\ 
\mu^2 \Bigl( \F 02_i - \frac N2\del_i K\Bigr) 
+ \frac 12 C_{ijk}& (x^j - \ii \mu\epsilon^j)(x^k - \ii \mu \epsilon^k) + 
\mu \Delta_{ij} (x^j - \ii \mu\epsilon^j)\biggr] \Psi = 0 \,.
\end{split}
\end{equation}
The remaining $\mu$-dependent terms can be removed by the substitution
\begin{equation}
\Psi \to \exp\Bigl[-\mu D_k\calt x^k - \mu \calt\lambda^{-1}
-\frac{\mu^2}{2} S^{jk} D_j\calt D_k\calt +\mu^2 S^k D_k\calt
\calt - \mu^2 S \calt^2- \mu^2 f^{(0,2)}\Bigr] \Psi
\end{equation}
where $S^{ij}$, $S^i$, and $S$ are the BCOV propagators,
defined up to holomorphic ambiguity by the equations
\begin{equation}
\begin{split}
\del_\ib S^{jk} &= \bar C_\ib^{jk} = \ee^{2K} G^{j\jb} G^{k\kb} \bar C_{\ib\jb\kb} \,, \\
\del_\ib S^j & = G_{\ib k} S^{jk} \,, \\
\del_\ib S &= G_{\ib k} S^k \,.
\end{split}
\end{equation}
To check that this substitution indeed removes the open string data, one has to use (among other things) the identity
\begin{equation}
\del_\lb\Bigl( \frac 12 C_{ijk} \epsilon^j\epsilon^k 
+ \ii \Delta_{ij} \epsilon^j + \frac 12\del_i\bigl(S^{jk} D_j\calt D_k\calt\bigr)
-\del_i \bigl(S^j D_j\calt \calt\bigr) + \del_i \bigl(S\calt^2\bigr)
\Bigr) = \ii \Delta_{ij} \bar\Delta_\lb^j \,,
\end{equation}
which provides the small phase space integration of the annulus anomaly (\cf,
\eqref{canbe}).

Summarizing these transformations, the shifted open topological string
partition function is
\begin{equation}
\eqlabel{look}
\begin{split}
\Psi^\nu (t^i,\bar t^\ib;x^i,\lambda^{-1}) &=  
\exp\Bigl[\mu D_k\calt x^k + \mu \calt\lambda^{-1} 
+\frac{\mu^2}{2} S^{jk} D_j\calt D_k\calt \\ -  \mu^2 S^k& D_k\calt
\calt + \mu^2 S \calt^2+ \mu^2 f^{(0,2)}\Bigr] \Psi(t^i,\bar t^{\bar i};x^i-\ii \mu\epsilon^i,
\lambda^{-1}+\ii \mu\epsilon) \,,
\end{split}
\end{equation}
with $\Psi$ as defined in \eqref{certain},
\begin{equation}
\eqlabel{certainagain}
\Psi(t^i,\bar t^\ib;x^i,\lambda^{-1})
= \lambda^{\frac{\chi}{24}-1- \mu^2 \frac N2}
\exp\biggl[ \sum_{\topa{g,h,n}{2g+h+n-2>0}}
\frac{\lambda^{2g+h+n-2}}{n!} \mu^h \F gh_{i_1,\ldots,i_n} x^{i_1}\cdots x^{i_n} 
\biggr] \,,
\end{equation}
in terms of the open-closed topological string amplitudes with D-brane
configuration given by the normal function, $\nu$. The statement is that
all $\Psi^\nu$ satisfy the same holomorphic anomaly equations, identical
to the anomaly equations satisfied by
the closed topological partition function $\Psi^{\rm closed} := \Psi^0$.

To see that the $\Psi^\nu$ are not all identical, it suffices to 
look at a few of the terms in the expansion of \eqref{look}. For
example, the leading behavior as $\lambda \to 0$ is given
by $\lambda^{\frac{\chi}{24}-1- \mu^2\frac N2}$, and thus depends on $N$.
More generally, because the sum in \eqref{certainagain} is restricted to 
$2g+h+n-2>0$, the leading terms in the $\lambda$-expansion are given
by the exponential prefactor in \eqref{look}, and hence depend 
explicitly on the brane configuration.

In terms of the diagrammatic expansion discussed in the previous subsection,
the fact that the shifted $\Psi^\nu$ all satisfy the same differential equation 
means that the Feynman rules, viewed as recursion relations between the $\F gh$, 
are independent of $\nu$. However, the different expansion around
$\lambda=0$ means that the initial conditions for the recursion relations
are different.  We stress that as a result the $\Psi^\nu$ are
not determined by $\Psi^0$ alone, {\it even before taking into account the 
holomorphic ambiguity}.

\section{Speculations}
\label{speculations}

We have seen above that any D-brane configuration specified by a normal 
function $\nu$ determines a solution $\Psi^\nu$ of the ordinary holomorphic
anomaly equation of BCOV.  This is achieved by forming the generating function
of open-closed topological string amplitudes, with a particular convention
for disk and annulus amplitudes with few insertions, and then shifting the
closed string variables in a certain way. This shift is different from
the one proposed in \cite{coy}. We have explained this statement in
various ways in both the small and large phase space. In this final section,
we present three applications of our results.

\subsection{Topological string wavefunctions}

Witten has proposed in \cite{wittenwf} that one should interpret the 
BCOV holomorphic anomaly equations as a statement of background independence
in the topological string. Although background independence does not
hold order by order in perturbation theory, the all-genus partition
function $\Psi(t^i,\bar t^\ib;x^i,\lambda^{-1})$, as a function of $x^i$,
$\lambda^{-1}$, can be thought of as a wavefunction defining a background
independent state in a certain auxiliary ``Hilbert space''. Witten's 
Hilbert space, $\calh_W$, arises by quantizing the symplectic
vector space $V_\reals = H^3(Y,\reals)$ in certain holomorphic polarizations, indexed
by the choice of background $(t^i, \bar t^\ib)$. The holomorphic anomaly
equations guarantee that as the background is varied, the wavefunction
$\Psi(t^i,\bar t^\ib;x^i,\lambda^{-1})$ varies precisely according to an 
infinitesimal Bogoliubov transformation, while the state $\vert \Psi \rangle 
\in \calh_W$ is background independent.

In this interpretation of the holomorphic anomaly equation, the fact that 
the shifted open string partition functions $\Psi^\nu$ also satisfy the 
ordinary BCOV equation simply means that they also define states in the 
same Hilbert space,
\begin{equation}
\lvert \Psi^\nu \rangle \in\calh_W \qquad \text{for all $\nu$}\,.
\end{equation}
The discussion in section \ref{discussion} shows that $\lvert \Psi^\nu \rangle$ 
indeed depends on $\nu$. We find it natural to conjecture that as $\nu$ varies 
over the set of all D-branes, the $\lvert \Psi^\nu \rangle$ will furnish a basis 
of $\calh_W$.

In support of this conjecture, we note that it is probably true at the
semi-classical level, \ie, at the level of the disk amplitude. 
Indeed, the topological disk partition function is given simply by the 
normal function itself. So we have to ask for the set of normal functions
that can be realized by wrapping D-branes, where for concreteness we
work in the context of the B-model on a Calabi-Yau $Y$. (Note that the usual 
definition of a normal function requires only Griffiths transversality, not that 
it be realizable algebraically.) It is by now well-accepted that the set of all
B-branes on $Y$ is equivalent to the derived category of coherent
sheaves, $D^b(Y)$. We obtain a normal function from any object in $D^b(Y)$
that deforms with $Y$ and whose second Chern class vanishes in $H^2(Y;\zet)$
\cite{griffiths}. We gave the typical example of this at the end of 
subsection \ref{specialgeometry}: Two homologous curves that deform with $Y$ 
define a normal function by integration over a bounding three-chain.

What is known mathematically is that the image of the so-called Griffiths
group ${\rm Griff}^2(Y)$ (homologically trivial algebraic cycles of 
co-dimension two, modulo algebraic equivalence) under the Abel-Jacobi map
to the intermediate Jacobian $J^3(Y)$ is not finitely generated as a vector 
space over $\rationals$ \cite{griffiths,infinite1,infinite2}. 
It appears likely\footnote{We thank D.~Morrison for conversations on this
issue.} that the image of $\griff^2(Y)$ in $J^3(Y)$ is in fact dense
(in the analytic topology). Lifting the period ambiguity in the definition
of a normal function, we conclude that the chain integrals \eqref{abeljacobi} 
will also be dense in $V_\complex/F^2 V_\complex\simeq T^*\tcalm$.

Now Witten's Hilbert space $\calh_W$ is a kind of quantization of the 
symplectic vector space $V_\reals$, in a complex polarization which identifies it with a
fiber of $T^* \tcalm$.  Although the exact relation is somewhat cumbersome (particularly
because the formal inner product in the complex polarization
is not positive definite), we view the denseness of the image
of the Abel-Jacobi map in $T^* \tcalm$ as evidence for the conjecture that 
the set of corresponding $|\Psi^\nu\rangle$ furnishes a (``Hilbert space'') 
basis of $\calh_W$.

\subsection{BPS state counting}

The wavefunction interpretation of the topological partition function plays
a crucial role in the formulation of the OSV conjecture \cite{osv}.  Namely,
consider the Type IIB superstring compactified on $Y$.  The resulting $d=4$
supergravity theory includes electrically and magnetically charged BPS states,
with charges $C \in V_\zet^*$.  OSV conjectured that the degeneracies
$\Omega(C)$ of these states are given by the Wigner function of $\vert \Psi \rangle$:
\begin{equation} \label{wigner}
\Omega(C) = \IP{\Psi| {\mathcal O}_C |\Psi}\,,
\end{equation}
where ${\mathcal O}_C$ is the Heisenberg group element associated to $C$.
To write \eqref{wigner} more concretely one has to choose a polarization.
For example, \cite{osv} uses the real polarization determined by a decomposition of the
$V_\reals$ into Lagrangian subspaces, $V_\reals = V_{electric} \oplus
V_{magnetic}$.  Then $\vert \Psi \rangle$ is represented by a function $\Psi(\chi)$, $\chi \in V_{magnetic}$,
the charge decomposes as $C = Q + P$, and \eqref{wigner} becomes
\begin{equation}
\Omega(Q,P) = \int d^n \chi \, \bar\Psi(\chi) \exp \left( Q^I \frac{\partial}{\partial \chi^I} + i P_I \chi^I \right) \Psi(\chi)\,.
\end{equation}

As we have seen, each D-brane configuration $\nu$ provides another state
$\vert \Psi^\nu \rangle$ in $\calh_W$; so one might ask what its Wigner function computes.
We are confused about various aspects of this question, but one possible guess follows.
Recall from \cite{extended} that $\nu$ corresponds to a brane configuration
with zero net charge.  The simplest example is a pair of homologous holomorphic curves $E_\pm$ in $Y$.
Consider wrapping a D3-brane of Type IIB on a 2-cycle in $Y$; this will give a string in the spatial $\reals^3$,
extended say in the $x^1$ direction.  This string supports an effective
field theory, with two supersymmetric vacua corresponding to wrapping the brane on the two curves $E_\pm$.
Now consider a ``domain wall'' configuration, i.e. Type IIB on $Y$ plus a brane which is in
the $E_-$ vacuum as $x^1 \to -\infty$ and the $E_+$ vacuum as $x^1 \to \infty$.  We regard this configuration 
as the ``background''.  Possible BPS states in this background include particles in the bulk of $\reals^3$
as well as ones bound to the domain wall on the string.  So a minimally invasive modification of the OSV conjecture 
is to propose that the Wigner function of $\vert \Psi^\nu \rangle$ counts such states.  (The quantization of charge in this sector is
non-standard --- there is a fractional part determined by $\nu$.  We do not understand how this affects the proposal.)

An open string extension of the OSV conjecture was also proposed in \cite{openbps}.  In
that case, however, one considers branes in a non-compact Calabi-Yau.  This leads to several salient
differences from the compact case:  the topological partition function
depends on the continuous moduli of the brane, and the physical theory includes additional BPS states
which couple to them.  As a result the proposal there took a somewhat different form.

\subsection{Hints from supersymmetry}

We have found that the holomorphic anomaly equations of the open and closed 
topological string can be transformed into one another by a simple shift of variables,
similar to that given in \cite{coy}.  This simple statement
deserves to have a physical interpretation.  In searching for one, however, we encounter an 
immediate puzzle:  the variables we shifted were not the physical moduli $(t^i, \bar t^{\bar{i}})$
of the worldsheet theory, but rather the formal generating-function parameters $(x^i, \lambda^{-1})$.
So to find a physical explanation of the shift we have to find a physical meaning for these 
parameters.

A possible clue comes from the fact that the equations and the shift take their simplest form when we trade the coordinates
$(t^i, \bar t^{\bar{i}}, x^i, \lambda^{-1})$ for $(X^I, y_I)$, \ie\ we use the complex structure of
$T^* \tcalm$, as was done in \cite{gnp}.  This complex manifold 
has a natural meaning from the spacetime point of view.  In the 
superconformal approach to $\caln = 2$ supergravity coupled to the vector multiplet moduli space $\calm$, one begins by 
considering a rigid (non-gravitational) theory with vector multiplet moduli space
$\tcalm$.  Upon classical dimensional reduction of this rigid theory from $d=4$ to $d=3$, say along $x^3$, supersymmetry dictates
that one obtains a theory with a hyperk\"ahler moduli space.  This space turns out to be (in one of its complex structures)
exactly $T^* \tcalm$  \cite{Cecotti:1988qn}.
So from this point of view the formal parameters $y_I$ become physical:  they represent the electric 
and magnetic Wilson lines $(A_I)_3, (A^D_I)_3$ of the $d=4$ gauge fields along the $x^3$ direction.
It would be very interesting to understand whether the shift relating the open and closed string 
can be related to turning on these gauge fields.  As support for this idea note that
$\lambda$ certainly is related to the graviphoton field strength \cite{bcov2}.

\begin{acknowledgments}
We would like to thank Juan Maldacena, Hirosi Ooguri, Martin Ro\v{c}ek, Cumrun Vafa, 
and Edward Witten for valuable discussions and communications. We are grateful to the
Aspen Center for Physics and to the Simons Workshop in Mathematics and Physics
for providing a stimulating atmosphere at some stages of this project.
The work of A.N.\ is supported in part by the Martin A.\ and Helen Chooljian
Membership at the Institute for Advanced Study, and by the NSF under
grant number PHY-0503584. The work of J.W.\ is supported in part by the
Roger Dashen Membership at IAS, and also by the NSF grant number PHY-0503584.
\end{acknowledgments}

\appendix

\section{From large to small phase space} \label{applargesmall}

In this appendix we give a few formulas which are needed for passing between 
the large and small phase space.

Recall the definition \eqref{we-defined},
\begin{equation}
X^I_i := \del_i X^I, \qquad X^I_{;i} := X^I_i + \del_i K X^I.
\end{equation} 
Tangent directions to $\calm$ and $\tcalm$ are related by
\begin{equation}
\frac{\del t^j}{\del X^J} X^J_i = \delta^j_i\,,
\quad
\frac{\del t^i}{\del X^J} X^I_{;i} = \delta_J^I - e^K \bar X^L 2\Im \tau_{LJ} X^I.
\end{equation}
The natural metric on $\tcalm$ is $\Im\tau$, related to 
the special K\"ahler data ($G$,$e^K$) on $\calm$ by
\begin{equation}
\eqlabel{metric}
e^{-K} = 2 \bar X^I X^J \Im\tau_{IJ}\,, \qquad
e^{-K} G_{i\bar j} = - X^I_{;i} \bar X^J_{;\bar j}\; 2\Im \tau_{IJ}\,, \qquad
\bar X^M 2\Im\tau_{MI} X^I_{;j}=0\,.
\end{equation}
The inverse metrics are similarly related by
\begin{equation}
\eqlabel{invmetric}
e^K G^{i\bar j} = - \frac{\del t^i}{\del X^I} \frac{\del \bar t^{\bar j}}{\del\bar X^J}
\; \frac 12 \Im \tau^{IJ}\,, \qquad
X^I_{;i} \bar X^J_{; \bar j} e^K G^{i\bar j} = - \frac 12 \Im\tau^{IJ} + e^K X^I \bar X^J\,.
\end{equation}
The connection and Yukawa coupling on $\calm$ are related to the 
data on $\tcalm$ by
\begin{equation}
\Gamma_{ij}^k - \del_i K \delta_j^k = 
\frac{\del t^k}{\del X^I} \del_i(X^I_{;j})
-\frac{\ii}{2} \frac{\del t^k}{\del X^I} X^M_{;i} X^K_{;j} C_{KM}^I\,,
\end{equation}
\begin{equation}
C_{ijk} = X^I_i X^J_j X^K_k C_{IJK} = X^I_{;i} X^J_{;j} X^K_{;k} C_{IJK}\,.
\end{equation}
where we have used the homogeneity \eqref{homoyuk} of the Yukawa coupling.
Using the foregoing relations one can verify in particular that $\Gamma_{ij}^k$ and $C_{ijk}$ have
the expected transformation properties.

Another useful fact, which may be verified using the last equation of \eqref{metric}, is
\begin{equation}
e^K 2  \bar X^M 2\Im\tau_{MI} \del_i(X^I_{;j}) x^j
=\frac \ii 2 X^M_i C_{KM}^I e^K \bar X^N 2\Im\tau_{NI} z^K\,.
\end{equation}
Finally, the infinitesimal invariants as defined in \eqref{delta-abs} and \eqref{delta-big}
are related by
\begin{equation}
\Delta_{ij} = X^I_i X^J_j \Delta_{IJ} = X^I_{;i} X^J_{;j} \Delta_{IJ}\,.
\end{equation}

\end{document}